\documentclass{PoS}
\usepackage[utf8]{inputenc}
\usepackage{amsmath}
\usepackage{bbm}
\usepackage[sort&compress]{natbib}
\bibpunct{[}{]}{,}{n}{}{}
\newcommand{\SA}{\ensuremath{\mathrm{S}}}
\newcommand{\Tr}{\ensuremath{\operatorname{Tr}}}
\newcommand{\SUN}{\ensuremath{\mathrm{SU}(N)}}
\newcommand{\U}{\ensuremath{\mathrm{U}(1)}}
\newcommand{\id}{\ensuremath{\mathbbm{1}}}      
\newcommand{\cpn}{\ensuremath{\text{CP}{(N-1)}}}
\newcommand{\pcm}{\ensuremath{\text{PCM}{(N)}}}
\newcommand{\cp}[1]{\ensuremath{\text{CP}{(#1)}}}
\renewcommand{\Re}{\ensuremath{\operatorname{Re}}}

\graphicspath{ {./plots/} {./} }

\title{A Lattice Study of Renormalons in Asymptotically Free Sigma Models}

\ShortTitle{A Lattice Study of Renormalons in Asymptotically Free Sigma Models}

\author{\speaker{Matthias Puhr} and Falk Bruckmann\\
        Institut für theoretische Physik, Universität Regensburg, \\ 93040 Regensburg, Germany\\
        E-mail: \email{matthias.puhr@ur.de}, \email{falk.bruckmann@ur.de}}

\abstract{
In general, perturbative expansions of observables in powers of the coupling constant in
quantum field theories are asymptotic series. In many cases it is possible to
apply resummation techniques to assign a unique finite value to an asymptotic series, but
a particular pattern of divergence, the so-called renormalon, gives rise to
non-perturbative ambiguities. The framework of Numerical Stochastic Perturbation Theory
(NSPT), based on stochastic quantisation and the perturbative expansion of lattice fields,
makes it possible to compute coefficients of perturbative series on the lattice. In this
work we report on an NSPT study of asymptotically free sigma models, namely the Principal
Chiral Model and the $\cpn$ model. We present results for a lattice computation of the
expansion coefficients of the energy density and discuss signatures of renormalons.
}

\FullConference{The 36th Annual International Symposium on Lattice Field Theory - LATTICE2018\\
		22-28 July, 2018\\
		Michigan State University, East Lansing, Michigan, USA.}

\begin{document}

\section{Introduction}
A heuristic argument suggest that perturbative expansions in powers of the coupling
constant~$\alpha$ in quantum field theories (QFTs) are generally asymptotic
series~\cite{Dyson:52:01}. To compare theoretical expectations with experimental results
it is necessary to assign a unique value to these asymptotic series. In many cases it is
possible to apply resummation techniques, like Borel summation, to unambiguously define
the value of an asymptotic series. Factorially divergent series with non-sign-alternating
coefficients, however, give rise to non-perturbative ambiguities ($\sim
\exp(-\frac{1}{\alpha})$) in the Borel summation. For the purpose of this work, we will
refer to this divergence pattern as renormalon, see~\cite{Beneke:99:01} for an in-depth
review of the topic.

The non-perturbative ambiguities introduced by renormalons raise fundamental questions
about the relation between perturbative expansions and a non-perturbative definition of
QFTs, which are adressed for example in resurgence theory (see~\cite{Unsal:16:01} for a
recent introductory review). Moreover, they can also play a crucial role for calculations in the
operator product expansion (OPE)~\cite{Beneke:99:01}. A better understanding of renormalon
physics is therefore of interest for a broad range of applications.

To clearly identify the divergence pattern giving rise to renormalons, one generally
has to calculate the perturbative expansion up to very high powers of $\alpha$. The number
of diagrams that have to be taken into account to calculate the coefficient at order
$\alpha^M$ in diagrammatic perturbation theory (DPT) is typically proportional to $M!$. For
most QFTs of interest it is in practice not feasible to go to large $M$. For this reason,
analytic verification of results obtained based on physical intuition or heuristic
arguments involving renormalons is often hard to achieve. An independent cross check of
renormalon based conjectures using lattice methods is therefore of great interest.

In this work we use the framework or Numerical Stochastic Perturbation Theory
(NSPT)~\mbox{\cite{DiRenzo:94:01,DiRenzo:04:01}} to study renormalons in sigma models in $d=1+1$
dimensions. These models are not trivial and have interesting features, like a
non-perturbative mass gap, confinement, or asymptotic freedom, in common with
QCD in $d=3+1$.  The big advantage of NSPT over DPT is that the numerical cost of
computing an expansion up to order $M$ is proportional to $M^2$ and not to $M!$, which enables
us to calculate perturbative expansions up to higher orders $M$ (state of the art
computations in $\mathrm{SU}(3)$ gauge theory in $d=3+1$~\cite{Bauer:12:01,Bali:13:01,Bali:14:02}
reach up to $M=35$).

\section{Numerical Setup}
We consider two different sigma models in $d=1+1$ dimensions, the Principal Chiral Model
($\pcm$) and the $\cpn$ model. Both models have been investigated in the context of
resurgence theory (see, e.g., ~\cite{Dunne:12:01,Dunne:13:01,Cherman:13:01,
Cherman:14:01}) and are expected to feature renormalons.  The lattice action of the $\pcm$ is
given by
\begin{equation}
  \label{eq:pcm_act}
  \SA = - 2 \beta N \sum\limits_{x,\nu}\Re\Tr\left(U(x)U({x+\nu})^\dagger  \right), \qquad
  U \in \mathbb{C}^{N\times N}, 
\end{equation}
where the fields $U$ sit on the sites of the lattice and are subject to the constraint
$UU^\dagger = \id$ and $\det U = 1$, i.e. $U \in \SUN$.
For the $\cpn$ model we use the lattice action
\begin{equation}
  \label{eq:cpn_aux_act}
\SA = - 2 \beta N \sum\limits_{x,\nu}\Re\left(n^\dagger(x)U_\nu(x)n(x+\nu)\right), \qquad n
\in \mathbb{C}^N,  
\end{equation}
where the $U_\nu \in \U$ are gauge links and the constraint on the fields is $n^\dagger n
= 1$. The gauge field is auxiliary, since there is no kinetic term for $U_\nu$ in
Equation~\eqref{eq:cpn_aux_act}.

The ansatz of NSPT is to expand the lattice fields in powers of $\alpha =
\beta^{-\frac{1}{2}}$ and truncate the series at a predefined power $M$,
so that, e.g., the fields of the $\pcm$ would be given by
\begin{equation}
  \label{eq:pcm_U_exp}
  U(x) = \sum\limits_{k=0}^M U_k(x)\beta^{-\frac{k}{2}}.  
\end{equation}
It is straightforward to define sums and products of truncated series:
\begin{equation}
\label{eq:sum_prod}
 U(x) + U'(x) =  \sum\limits_{k=0}^M (U_k(x) + U'_k(x)) \beta^{-\frac{k}{2}} , \qquad  U(x) \cdot U'(x) =
      \sum\limits_{k=0}^M \left(\sum\limits_{l=0}^k U_l(x)U'_{l-k}(x)\right) \beta^{-\frac{k}{2}}.
\end{equation}
Analytic functions of the field variables can be defined by simply inserting the expansion
in the Taylor series of the function and neglecting terms of order
$\mathcal{O}(\beta^{-\frac{M+1}{2}})$. The numerical cost of computing the product in
Equation~\eqref{eq:sum_prod} scales like $M^2$ and since this is the most expensive
operation we expect the cost of NSPT to be also proportional to $M^2$.\footnote{This
estimate ignores possible bottlenecks like cache sizes or memory bandwidth.}

Monte Carlo algorithms with an accept/reject step can not be expanded in powers of
$\beta^{-\frac{1}{2}}$ and are not compatible with expansions of the
form~\eqref{eq:pcm_U_exp}. NSPT therefore employs the discretised Langevin equation to
generate lattice configurations~\cite{Batrouni:85:01}. Discretising the Langevin time
introduces a systematic error and makes it necessary to simulate at many different
discrete Langevin time steps $\epsilon$ and to perform an extrapolation $\epsilon \to 0$. The
size of the systematic error depends on the integration routine used in the discretised
Langevin update. We use the second order Runge-Kutta type integrator
from~\cite{Bali:13:01}, for which the systematic error is of order
$\mathcal{O}(\epsilon^2)$.

For this first exploratory study we consider the perturbative expansion of the energy
density, which is given by  
\begin{equation}
  \label{eq:energy_density_def}
  E_\text{CPN} = 2 N \langle \Re\left(n^\dagger(x)U_\nu(x)n(x+\nu)\right) \rangle , \qquad
  E_\text{PCM} = 1 - \frac{1}{N}\langle \Re\Tr\left(U(x)U({x+\nu})^\dagger\right)  \rangle 
\end{equation}
for the $\cpn$ model and $\pcm$, respectively. The perturbative series for the energy density
is formally given by
\begin{equation}
  \label{eq:energy_density_pert}
  E = \sum\limits_{n} E_{n} \beta^{-\frac{n}{2}}, 
\end{equation}
but the coefficients with odd $n$ vanish identically. (To see why remember that the free
energy, up to shifts and rescaling, is given by partial derivative of the partition
function with respect to $\beta$.) One reason for studying the energy density is that
analytic results for the lowest expansion coefficients in~\eqref{eq:energy_density_pert}
are available in the literature for both models considered in this work.

\section{Results}
\begin{figure}[h]
  \centering
  \includegraphics[width=0.49\textwidth]{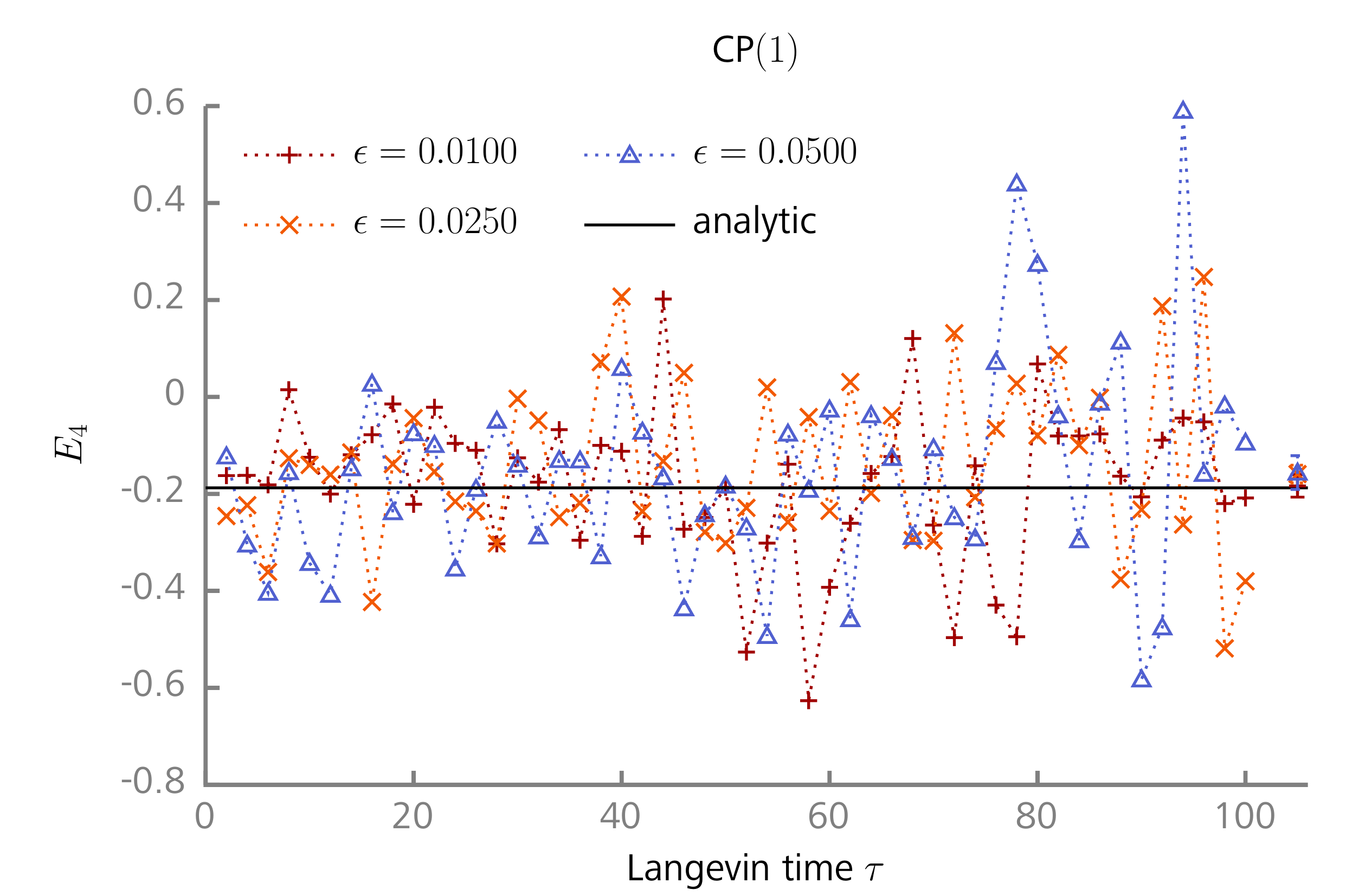}
  \includegraphics[width=0.49\textwidth]{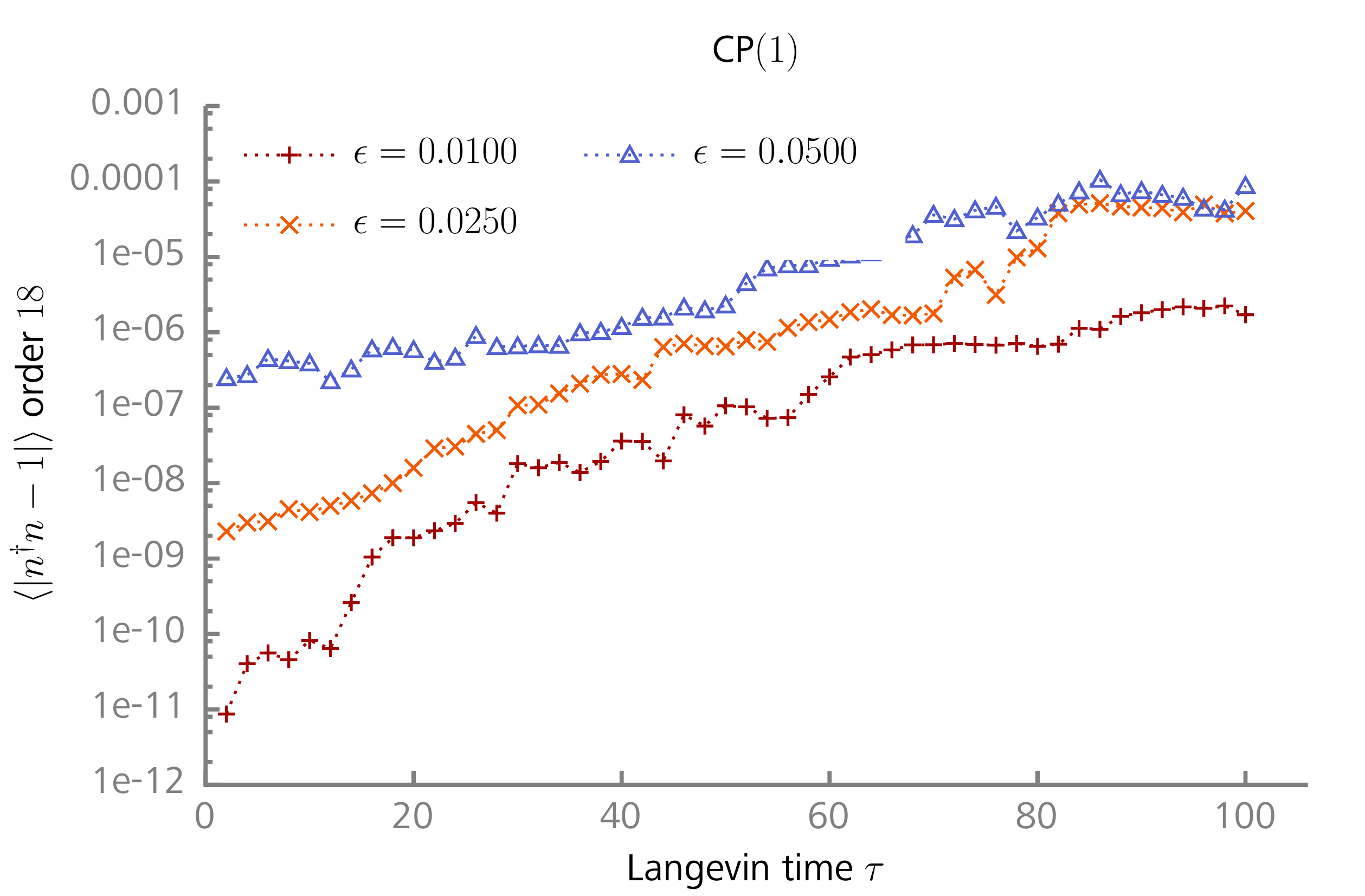}  
  \caption{Left: Langevin trajectories for the coefficient $E_4$ in the $\cp{1}$ model on
    a $V = 16 \times 16$ lattice for different values of $\epsilon$. The symbols with
    error bars on the very right show the average over the trajectory and the solid black
    line is the continuum perturbation theory result. Right: Average deviation from the constraint
    $n^\dagger n = 1$ at order $18$ as a function of the Langevin time $\tau$. All
    parameters are the same as for the left plot.}
  \label{fig:cpn_res}
\end{figure}

We start this section with a discussion of the results for medium sized lattices with
$V=16 \times 16$. To check the dependence on the discrete Langevin time step we performed
NSPT calculations for $\epsilon = 0.05, \ 0.025$ and $0.01$. For the $\cpn$ model the
first three non-trivial expansion coefficients are known and we find good agreement between our
results and the analytical values, even for the largest $\epsilon$. A typical Langevin
history for $E_4$ in the $\cp{1}$ model is shown on the left plot in
Figure~\ref{fig:cpn_res}. While the results for the first few coefficients are very
encouraging, there seems to be a problem with higher order coefficients. We find that the
constraint $n^\dagger n = 1$, which has to be fulfilled order by order, is violated for
higher orders at large Langevin times. The reason seems to be an accumulation of numerical
errors over the Langevin run\footnotemark. As an example, we plot the deviation from the constraint at
order $18$ on the right hand side of Figure~\ref{fig:cpn_res}.
\footnotetext{We use a formulation of the Langevin algorithm which, for infinite precision
  arithmetic, exactly preserves the constraint even for finite $\epsilon$.}

\begin{figure}[h]
  \centering
  \includegraphics[width=0.49\textwidth]{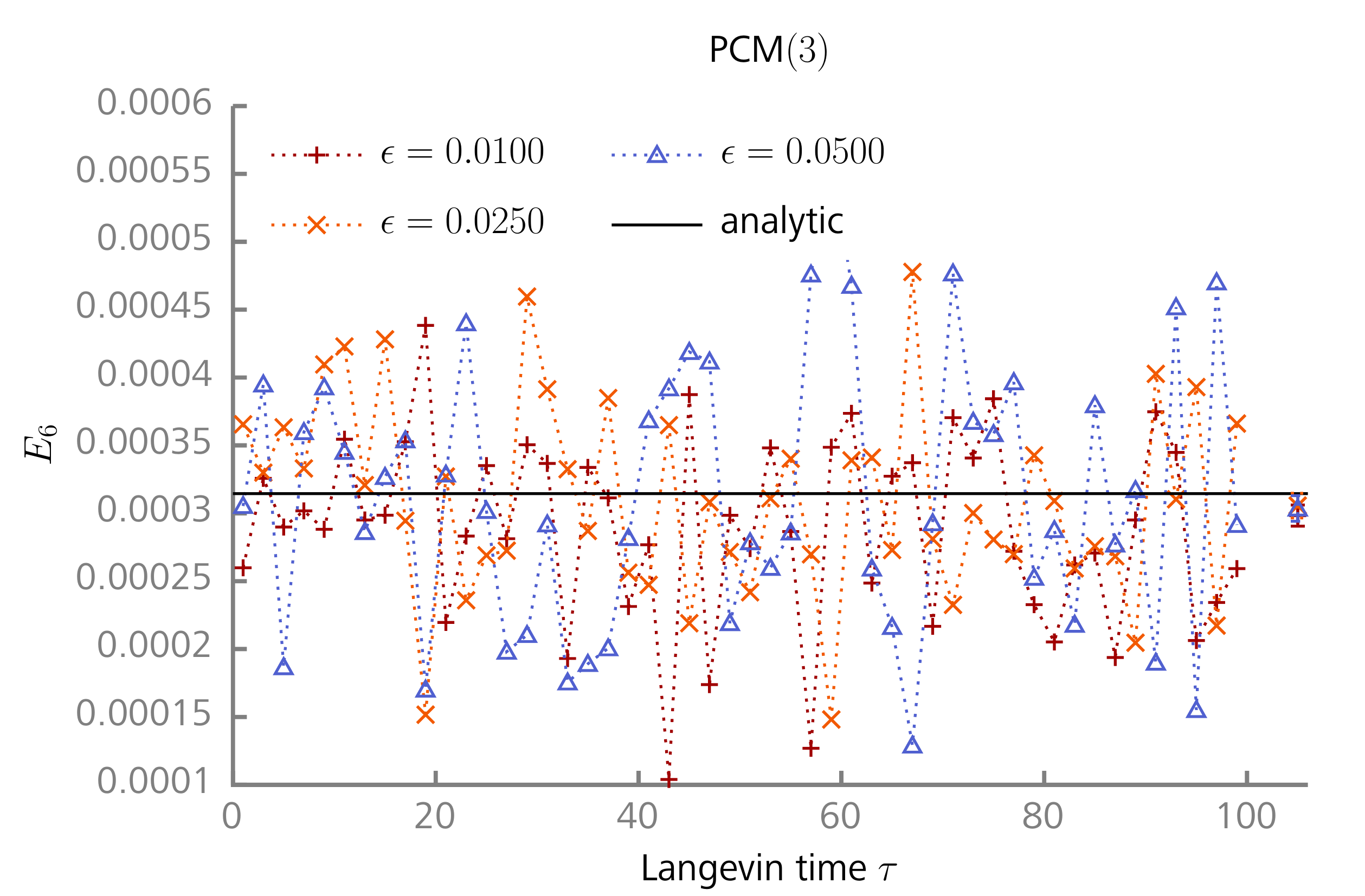}
  \includegraphics[width=0.49\textwidth]{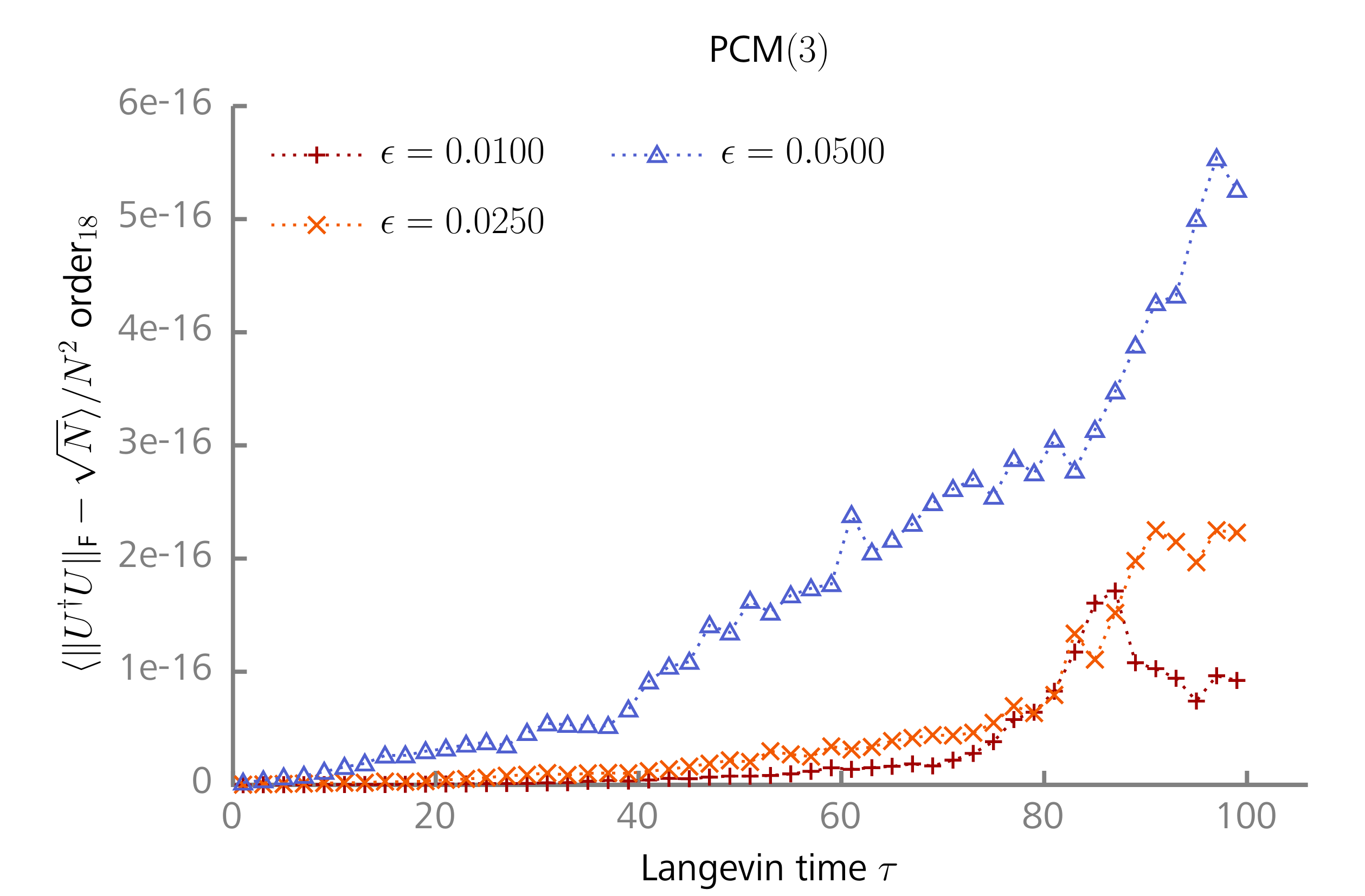}
  \caption{Left: Langevin trajectories for the coefficient $E_6$ in $\text{PCM}(3)$ on
    a $V = 16 \times 16$ lattice for different values of $\epsilon$. The black
    line marks the continuum perturbation theory result and the symbols with
    error bars on the very right show the average over the trajectory. Right: Average
    deviation from the constraint    $ U^\dagger U = \id$ at order $18$ as a function of
    the Langevin time $\tau$. The deviation from the unit matrix is measured as $|
    \|U^\dagger U\|_F - \sqrt{N}|/N^2$, where $\|\|_F$ stands for the Frobenius norm. All
    parameters are the same as for the left plot.}
  \label{fig:pcm_res}
\end{figure}

For the $\pcm$ the coefficients are known up to $E_6$. Again,  we generally find a good
agreement between the analytic values and the NSPT calculations, see the left plot in
Figure~\ref{fig:pcm_res}. We do not observe any large deviations from the constraint
$U^\dagger U = \id$: as the plot on the right panel of Figure~\ref{fig:pcm_res} shows, the
constraints are fulfilled even at large orders (up to some small errors of the order of
the machine epsilon).

\begin{figure}[h]
  \centering
  \includegraphics[width=0.45\textwidth]{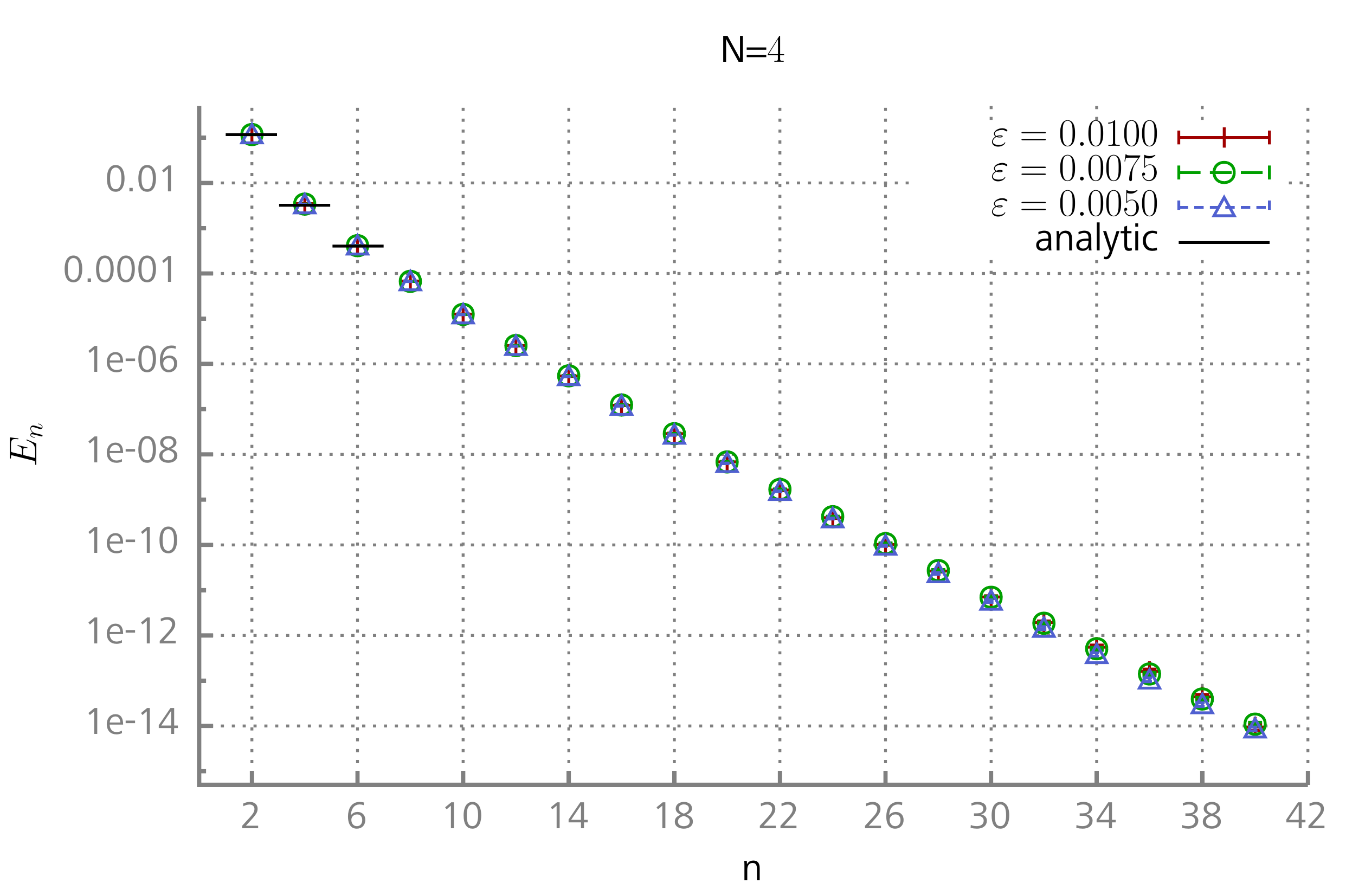}
  \includegraphics[width=0.45\textwidth]{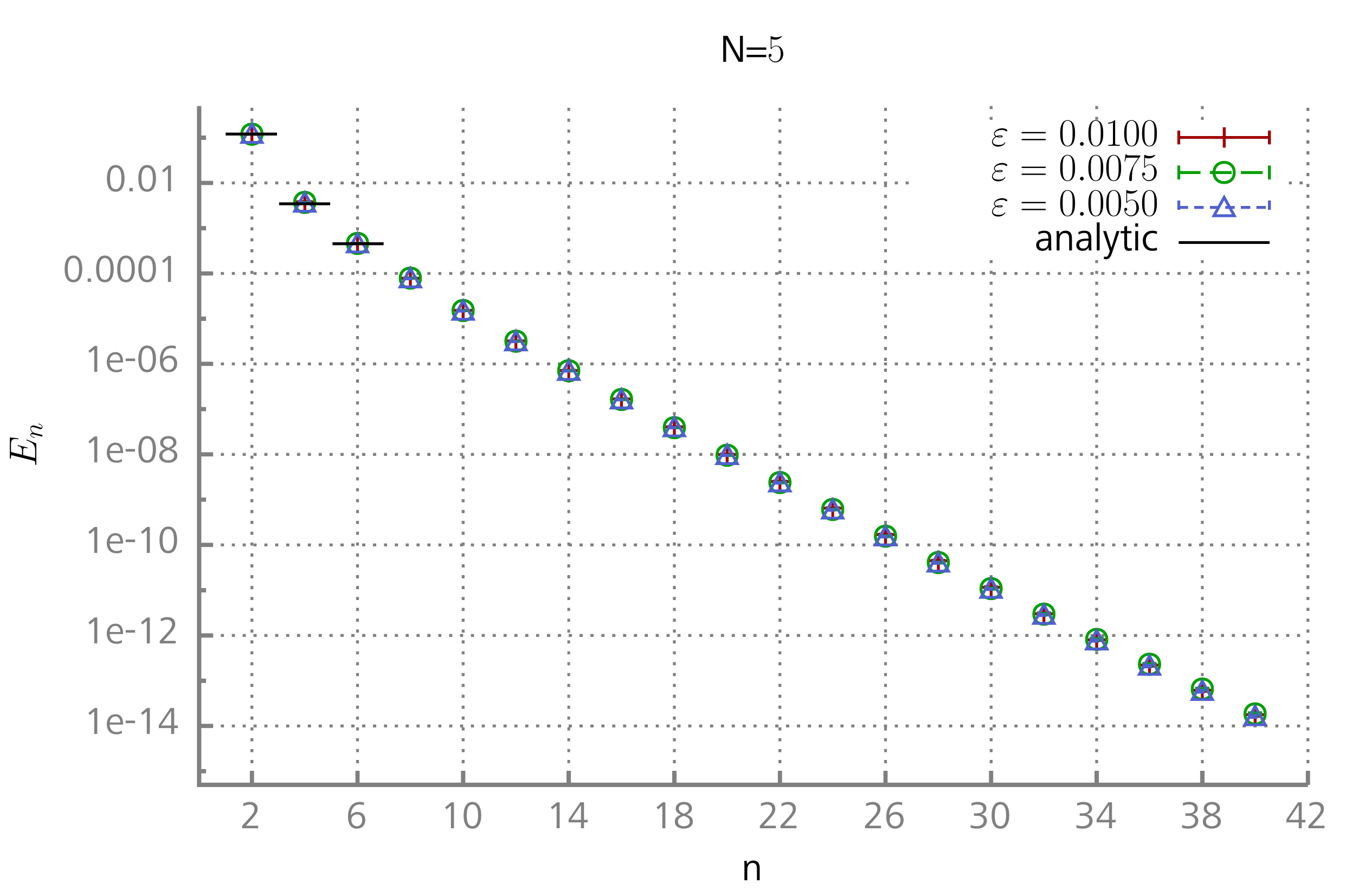}
  \includegraphics[width=0.45\textwidth]{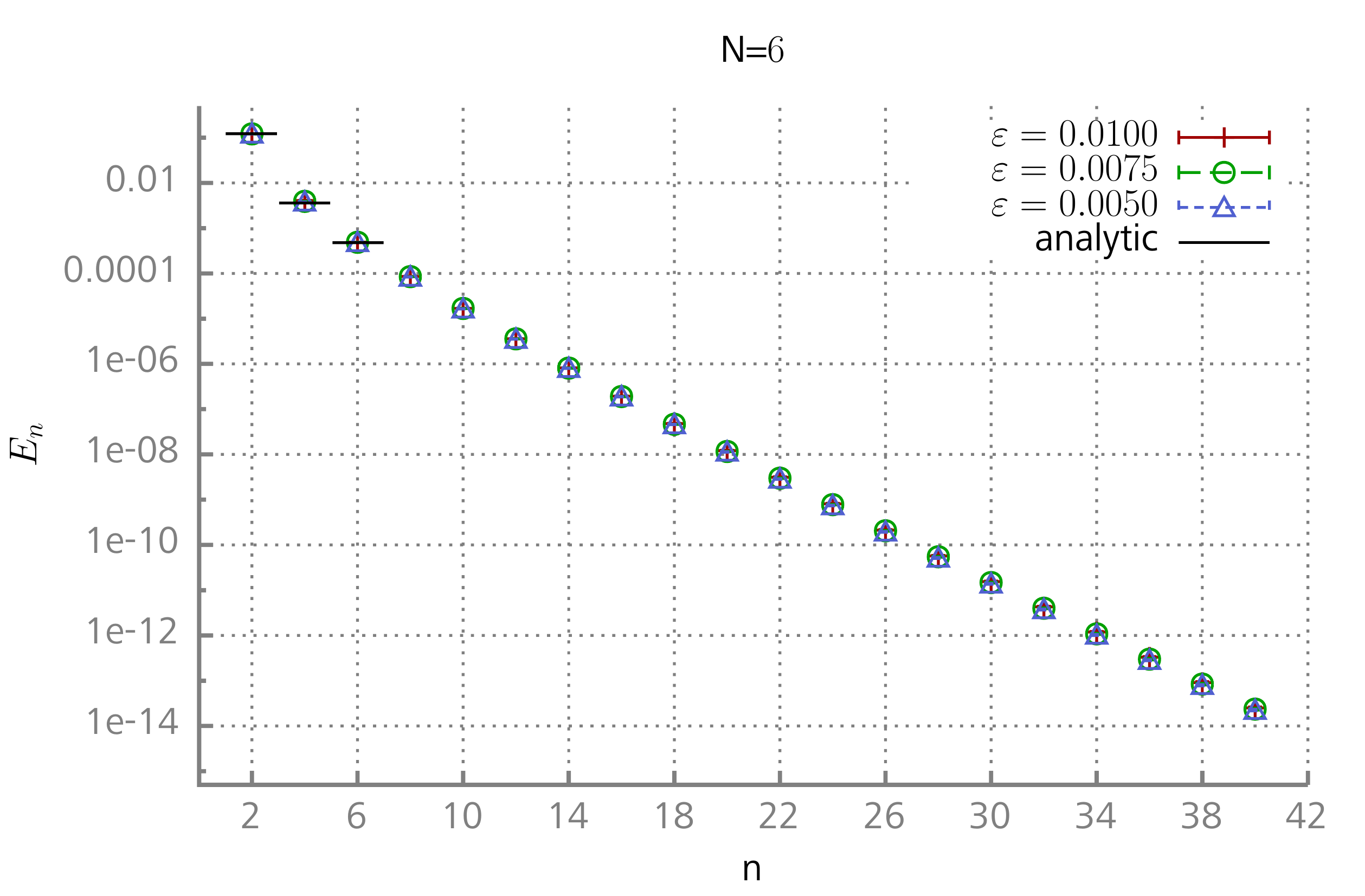}
  \includegraphics[width=0.45\textwidth]{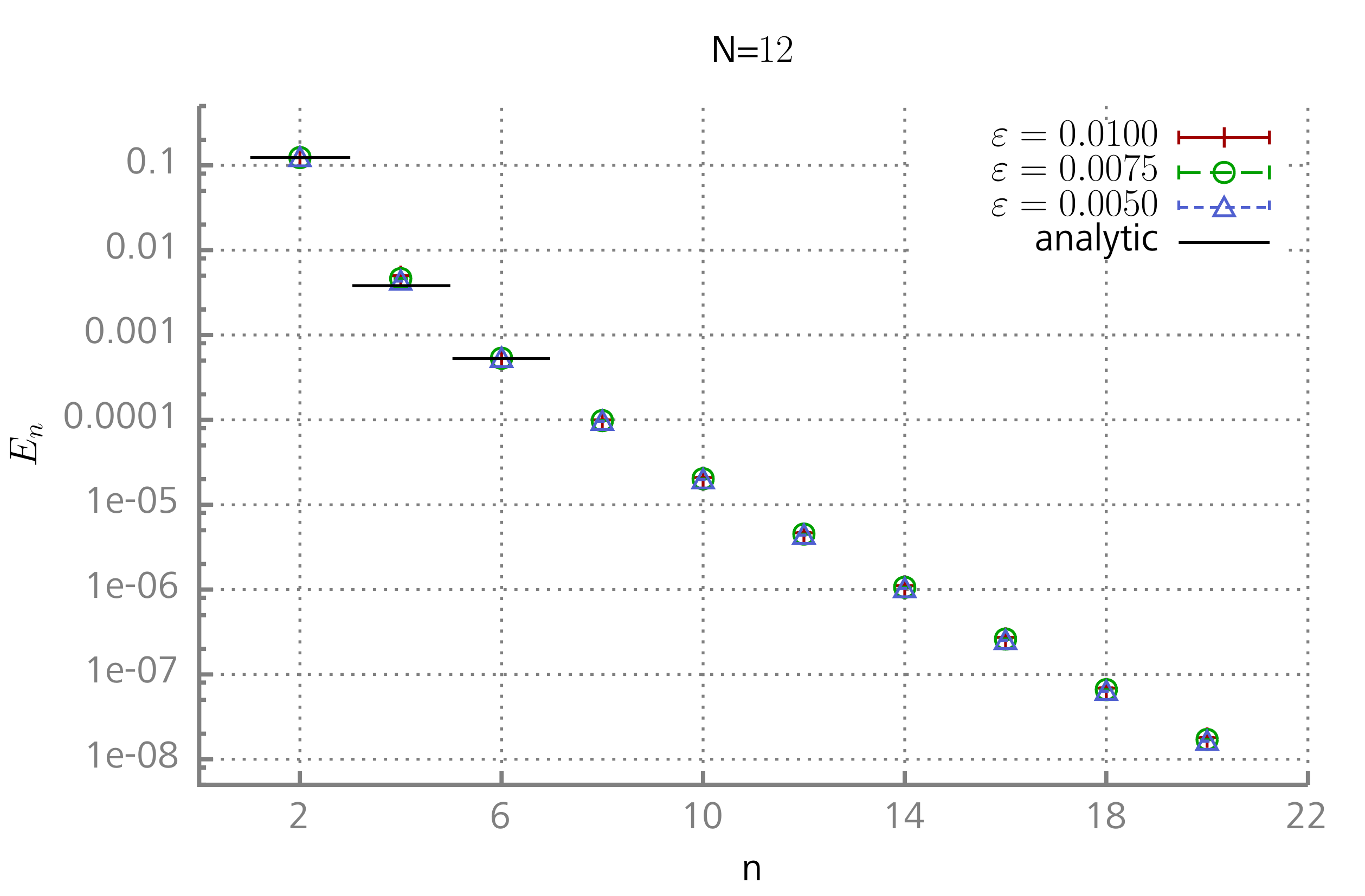}
  \caption{Results for the $\pcm$ on a lattice of size $V = 32 \times 32$ for different
    values of $N$ and $\epsilon$. The plots show the (even)
    expansion coefficients. Where available the analytical results, denoted by a small
    black line, are plotted for comparison.  }
  \label{fig:pcm_largeV}
\end{figure}

Simulations on a larger lattice with $V = 32 \times 32 $  were only performed for the $\pcm$,
because of the constraint compliance issues with the $\cpn$ model. On the larger lattice
we simulate the $\pcm$ for several different values of $N=4,5,6,12$ and use  perturbative
expansions up to order~$40$ (order~$20$ for $N=12$) in $\beta^{-\frac{1}{2}}$.  Moreover, we
significantly decrease the size of the Langevin time steps and use $\epsilon =
0.01, 0.0075,0.005$. The results are summarised in Figure~\ref{fig:pcm_largeV} and show
that NSPT enables us to calculate perturbative expansion coefficients with high precision
up to the highest orders considered in this work.  

If the expansion~\eqref{eq:energy_density_pert} is divergent with a renormalon pattern,
the expectation is that asymptotically for large $n$ the coefficients have the form $E_n
\sim b^n n! n^c$, with constants $b,c$ and $b>0$. Assuming the coefficients indeed
have this asymptotic behaviour, up to an optimal truncation order $n^*$ the terms $E_n
\beta^{-\frac{n}{2}}$ decrease and the inclusion of additional terms in
the series gives a better approximation to $E$. It can be shown that $n^* \sim
\frac{\beta^\frac{1}{2}}{b}$ and OPE arguments suggest that, to leading order,
$b \sim \frac{\beta_0}{D}$, where $\beta_0$ is the leading coefficient in the expansion of
the $\beta$-function and $D$ is the energy dimension of the observable in question ($D=2$
for E). For the $\pcm$ with action~\eqref{eq:pcm_act} $\beta_0 = \frac{1}{8\pi}$.
The minimal coefficient $E_n$ can be derived by the same arguments to occur at $n^* \sim 16 \pi$,
which is larger than our maximal expansion order $M=40$. We would have to go to much
higher orders before the expansion coefficients start to increase and the renormalon
pattern emerges.

\begin{figure}[h]
  \centering
   \includegraphics[width=0.65\textwidth]{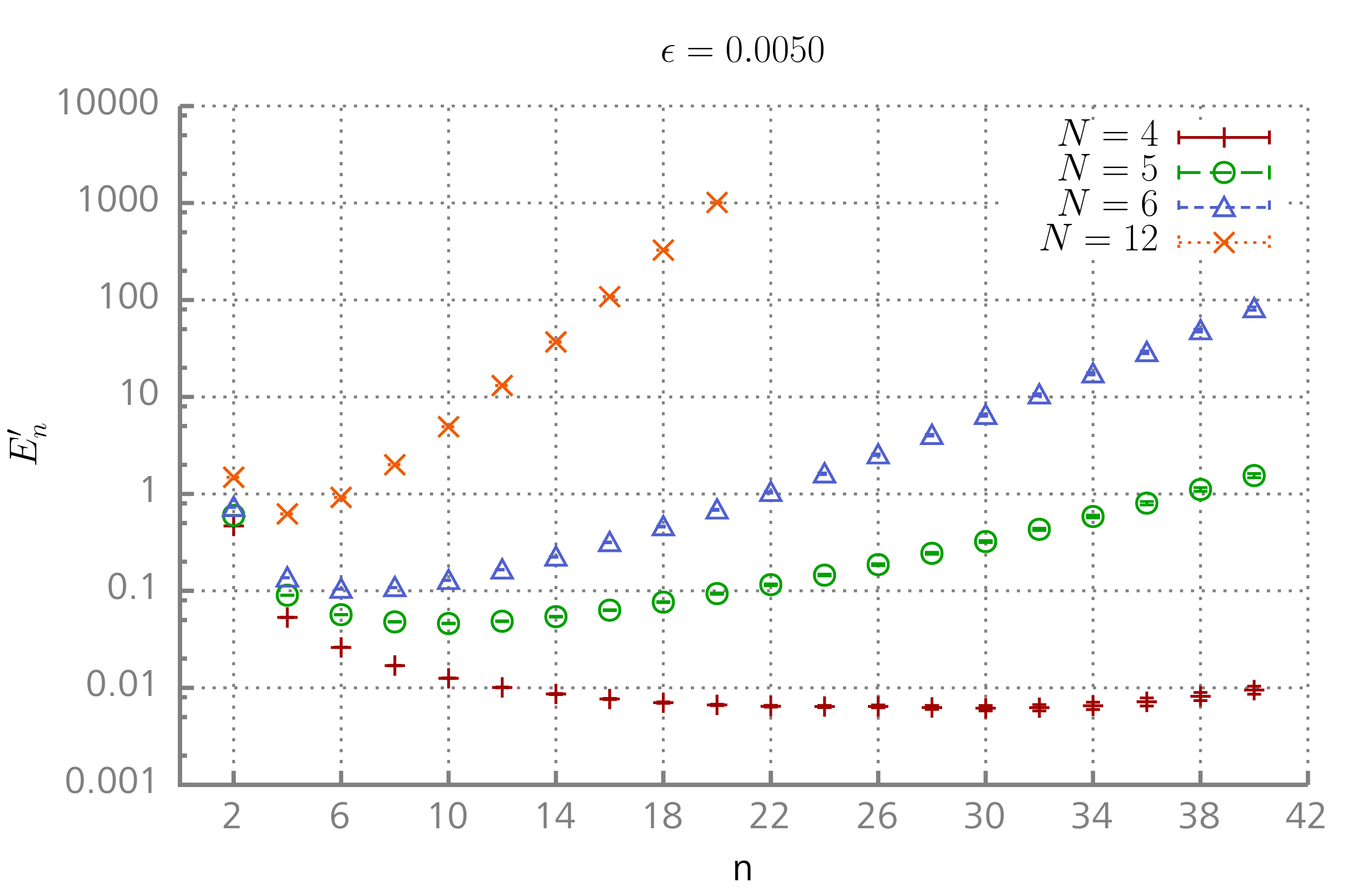}
  \caption{Expansion coefficients of the energy density $E'_n$ in the $\pcm$ for
    rescaled coupling $N \beta =: \beta'$. The Langevin time step is $\epsilon=0.005$ for
    all $N$ in this plot.
  }
  \label{fig:res_coef}
\end{figure}

Fortunately there is an alternative, less expensive way to check for renormalons.
Rescaling the coupling $\beta' := \beta N$ leads to $\beta_0' = \frac{N}{8 \pi}$ and gives
an $N$-dependent $n'^* \sim \frac{16 \pi}{N}$. There is no need to recalculate the
expansion coefficients, since the new ones are also given by a simple rescaling: $E'_n =
E_n N^{\frac{n}{2}}$.  Figure~\ref{fig:res_coef} shows a plot of the rescaled coefficients
$E'_n$. With the exception of $N=4$ the expectation for the optimal truncation order
($n'^* \approx 13,10,8,4$ for $N=4,5,6,12$) is in good agreement with a change of the
monotonicity behaviour of the expansion coefficients. 

\section{Discussion and Outlook}
In this work we applied NSPT to calculate the expansion coefficients of the energy
density in the $\pcm$ and the $\cpn$ model. For the $\cpn$ model we found that we can compute
the lowest expansion coefficients with high precision, but for the higher coefficients the
results are spoiled by large deviations from the constraint $n^\dagger n = 1$. We tried to
to fix this by going to smaller time steps $\epsilon$, periodically rescaling the fields
to enforce the constraint ``by hand'', and even by considering a different lattice
discretisation of the $\cpn$ action\footnote{The equations of motion can be used to get
rid of the auxiliary gauge field, giving the so-called ``quartic action''.}. So far, none
of these approaches has been entirely successful and the constraint violation remains an
open problem.

For the $\pcm$ our numerics worked very well and we were able to calculate expansion
coefficients up to order $40$. Even for the largest orders constraint violation does not
pose a problem in the $\pcm$ calculations. Our results for the monotoneicity behaviour of the
rescaled coefficients, shown in Figure~\ref{fig:res_coef}, are in general in good
agreement with expectations and can be interpreted as renormalon signatures.

Further work is necessary to unambiguously identify a renormalon in the $\pcm$.
The constant $b$ can be extracted from ratios of expansion coefficients and should be a
more reliable indicator of a renormalon. To this end, a careful execution of the
extrapolation $\epsilon \to 0$ and the limit $V \to \infty$ is
necessary~\cite{Bali:13:01,Bali:14:02}. This is currently work in progress and the final
results will be published elsewhere.

\emph{Acknowledgements:} We thank G.~Bali, P.~Buividovich, G.~Dunne and M.~Ünsal for
helpful discussions. Large parts of our numerical code were developed together with
J.~Simeth. We acknowledge support from DFG (Contract No.\ BR 2872/6-2, BR 2872/8-1).
The computations were performed on a local cluster at the University of Regensburg
and on the Linux cluster of the LRZ in Garching.



\providecommand{\href}[2]{#2}\begingroup\raggedright\endgroup

\end{document}